\documentclass[apl,amsmath,amssymb,reprint]{revtex4-1}
\usepackage{graphicx}
\usepackage{dcolumn}
\usepackage{bm}
\newcommand{\p}{$\%$}

\newcommand{\pn}{$\mathrm{R{_{N_2}}}$}

\newcommand{\GMM}{$\alpha^{\prime\prime}\mathrm{-Fe_{16}N_{2}}$}

\newcommand{\tfn}{$\mathrm{Fe_{4}N}$}
\newcommand{\tcn}{$\mathrm{Co_{4}N}$}

\newcommand{\tmn}{$\mathrm{TM_{4}N}$}
\newcommand{\muB}{$\mathrm{\mu_{B}}$}

\begin{document}

\title{Phase formation, thermal stability and magnetic moment of cobalt nitride thin films}

\author{Rachana Gupta$^a$} \author{Nidhi Pandey$^b$}\author{Akhil Tayal$^b$} \author{Mukul Gupta$^b$} \email{mgupta@csr.res.in/dr.mukul.gupta@gmail.com}
\address{$^a$Institute of Engineering and Technology DAVV, Khandwa Road, Indore 452 017,
India\\$^b$UGC-DAE Consortium for Scientific Research, University
Campus, Khandwa Road, Indore 452 001, India}


\date{\today}


\begin{abstract}

Cobalt nitride (Co-N) thin films prepared using a reactive
magnetron sputtering process by varying the relative nitrogen gas
flow (\pn) are studied in this work. As \pn~increases, Co(N),
\tcn, Co$_3$N and CoN phases are formed. An incremental increase
in \pn, after emergence of \tcn~phase at \pn=10\p, results in a
continuous expansion in the lattice constant ($a$) of \tcn. For
\pn=30\p, $a$ maximizes and becomes comparable to its theoretical
value. An expansion in $a$ of \tcn, results in an enhancement of
magnetic moment, to the extent that it becomes even larger than
pure Co. Though such higher (than pure metal) magnetic moment for
Fe$_4$N thin films have been theoretically predicted and evidenced
experimentally, higher (than pure Co) magnetic moment are
evidenced in this work and explained in terms of large-volume
high-moment model for tetra metal nitrides.

\end{abstract}

\maketitle

Nitrides of 3$d$ magnetic transition metals (TM=Cr,Mn,Fe,Co,Ni)
are an interesting class of materials for applications in magnetic
devices. With inclusion N atoms ($\sim10-20~at.\%$), TM become
chemically inert and since they preserve the metallic character of
the host metal, their structural and magnetic properties are
superior. For example, a higher-than-Fe magnetic moment for iron
nitrides (\GMM~\cite{APL11_Ji} and
\tfn~\cite{APL11_Ito_Fe4N,APL08_Atiq}), has been the driving force
for the intense research work in this
system.~\cite{JPCC15_Bhatta,PRB14_MG} Specially, in case of tetra
TM nitrides (\tmn), such enhanced magnetic moment is caused by a
volume expansion (compared to a hypothetical $fcc$
metal).~\cite{JAC14_Imai} All \tmn~share a common $fcc$ structure,
in which metal atoms are arranged in the $fcc$ positions and N
atoms occupy the body centered positions. Such incorporation of N
atoms results in an expansion of the $fcc$
lattice.~\cite{PRB07_Matar}

Compared to the Fe-N system, the Co-N system has been relatively
less
explored.~\cite{APL86_Matsuoka,JMS87_Oda,JAC95_Suzuki,Vac01_Asahara,JVSTA:Fang:CoN,MSEB08_Jia,SSC08_Paduani,JMMM10_Matar,JMMM10_Imai}
Recent theoretical calculations predicted that the spin
polarization ratio for \tcn~is even higher than that of
\tfn.~\cite{JMMM11_Takahashi} This has lead to somewhat renowned
interests in the Co-N system both theoretically and
experimentally.~\cite{APL11_Ito_Co4N,JCS11_Ito,ECS12_Bhandari,JAC14_Lourenco,JAC14_Liu,TSF14_Silva,JAC15_Silva}
Though theoretical studies predict that under large-volume
high-moment approach, the magnetic moment of \tcn~can be larger
than Co,~\cite{PRB07_Matar} experimental results always find a
value much smaller than pure Co, for \tcn~ thin
films.~\cite{APL11_Ito_Co4N} In this Letter we report more than Co
magnetic moment for \tcn~thin films.


Co-N thin film samples (200\,nm thick) were deposited on glass
substrate at room temperature using a direct current magnetron
sputtering (dcMS) process operating at constant power of 50\,W. A
Co(purity 99.95\%) target (75\,mm diameter) was sputtered using a
mixture of N$_2$(99.9995\%) and Ar(99.9995\%) gases. Relative
nitrogen gas flow defined as \pn =
$\mathrm{p{_{N_2}}}$/($\mathrm{p{_{N_2}}}$+$\mathrm{p_{Ar}}$)$\times$100
(where $\mathrm{p{_{N_2}}}$ is N$_2$ and $\mathrm{p_{Ar}}$ is Ar
gas flow), was varied at 0, 5, 10, 20, 30, 50, 75 and 100\%. The
total gas flow was fixed at 10\,sccm. With a base pressure of
1$\times10^{-7}$\,mbar, the pressure during deposition was about
3$\times10^{-3}$\,mbar. The dcMS system was suitably modified to
deposit all samples sequentially on a 25\,cm long substrate kept
at a distance of 7\,cm from the target. The magnetron source was
masked with a 10\,mm wide slit. After the deposition for a \pn,
substrate was moved linearly for deposition of next sample. X-ray
reflectivity (XRR) and diffraction (XRD) measurements were carried
out using laboratory x-ray systems equipped with a Cu k-$\alpha$
x-rays. Thermal stability was studied after successive thermal
annealing and XRD measurements.



\begin{figure}\center
\includegraphics [width=70mm,height=55mm] {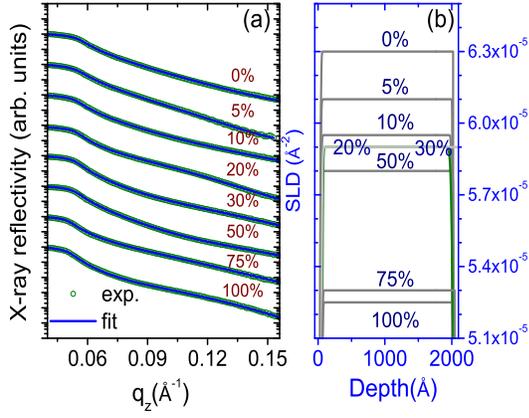}
\caption{\label{fig:xrr} (Color online) XRR pattern of Co-N thin
films deposited for various \pn (a) and obtained scattering length
densities from fitting of XRR data (b). XRR patterns are shifted
vertically, for clarity.} \vspace{-5mm}
\end{figure}

XRR patterns of Co-N thin films deposited for various \pn~are
shown in fig.~\ref{fig:xrr}(a), they were fitted using Parratt's
formulism.~\cite{Parratt32} Though total thickness oscillations
could not be seen in XRR pattern due to large thickness, still a
vital information about the film density is obtained, which
decreases gradually as \pn~is increased. Obtained values of
scattering length densities are plotted in fig.~\ref{fig:xrr} (b).
Roughness of film surface was typically about 1.3-1.5\,nm.

\begin{figure}\center
\includegraphics [width=70mm,height=80mm] {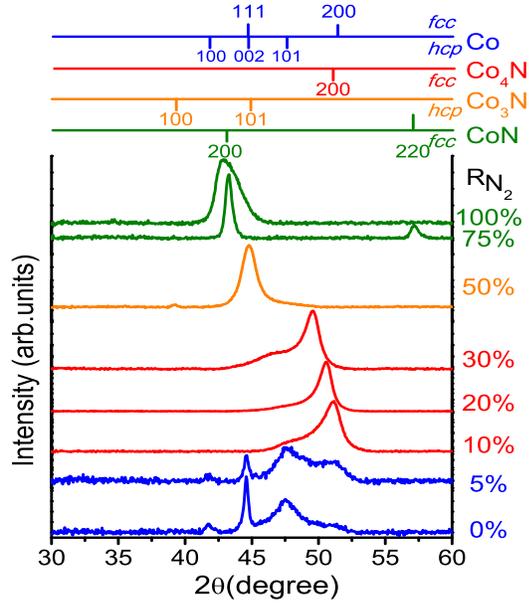}
\caption{\label{fig:xrd} (Color online) XRD pattern of Co-N thin
films deposited for various \pn.} \vspace{-5mm}
\end{figure}

XRD pattern of samples deposited for different \pn~are shown in
fig.~\ref{fig:xrd}. For \pn=0\p, pure Co $hcp$ phase is observed
however a faint peak at 2$\theta$=51.4 degree corresponding to
$fcc$ (200) reflection can also be seen. Co is known to stabilize
in $hcp$ phase below 690\,K and above it, in $fcc$
phase.~\cite{JVST99_Zhang,PRL00_newCo} However, co-existence of
both phases in thin films is also seen.\cite{MCP03_Ko,RPP08_Vaz}
For the sample prepared at \pn=5\p, the structure remains similar
to 0\p~sample, but peak widths become broader due to interstitial
incorporation of N atoms.~\cite{Gupta:PRB05} As \pn~increases to
10\p, the structure changes and reflection corresponding to
\tcn(200) can be seen. A rather broad tail on the onset of this
peak, is due to \tcn(111) (discussed later). An increase in
\pn~for 20 and 30\p, leads to shift in this peak towards lower
angles indicating an expansion in $a$ of \tcn(see
table~\ref{table1}). For \pn = 50\p, the phase identified is $hcp$
Co$_3$N and for \pn=75\p, the structure changes again to $fcc$
cobalt mononitride(CoN). For \pn=100\p, CoN(200) peak gets
broadened due to nanocrystallization.~\cite{Gupta:PRB05} To
correlate N at.\% with the structure of samples, secondary ion
mass spectroscopy measurements were performed. Using a procedure
adopted for Fe-N thin films,\cite{MG:JAC:2011,gupta:JAP2011} we
find N at.\% for \pn=30 and 100\% samples is 20$\pm$3 and
50$\pm$4, respectively, which is expected for \tcn and CoN
structures. Grain sizes and $a$ obtained from the most intense
peak in XRD pattern are given in table~\ref{table1}.

\begin{figure}\center
\includegraphics [width=85mm,height=65mm] {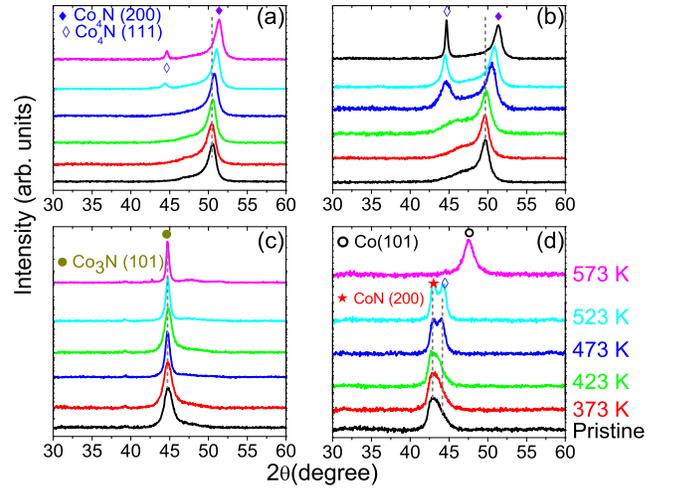}
\caption{\label{fig:ann} (Color online) XRD pattern of Co-N thin
films in the pristine state and after annealing them at different
temperatures for \pn=20\%(a), 30\%(b), 50\%(c) and 100\%(d).}
\vspace{-5mm}
\end{figure}

\begin{table} [!hb] \vspace{-5mm}
\caption{\label{table1} Parameters, crystal structure, grain size,
lattice constant ($a$) and average magnetic moment ($\mu$) per Co
atom for Co-N thin film samples prepared for various \pn. For
samples having $hcp$ structure $a$=c is taken and cell parameter
a=b=c/1.62.}
\begin{ruledtabular}
\begin{tabular}{lcccc}
\pn&crystal&grain&$a$&magnetic\\
(\p)&structure&size(nm)&(nm)&moment($\mu_\mathrm{B}$)\\
&($\pm$0.05)&($\pm$1)&($\pm0.004$)&($\pm$0.05)\\\hline
0&$hcp+fcc$&33&0.407&1.7\\
5&$hcp+fcc$&22&0.406&1.6\\
10&$fcc$&6&0.358&1.6\\
20&$fcc$&9&0.362&1.75\\
30&$fcc$&7&0.370&1.85\\
50&$hcp$&7&0.431&1.5\\
75&$fcc$&19&0.418&0\\
100&$fcc$&10&0.423&0\\
\end{tabular}
\end{ruledtabular}
\end{table}

To resolve the structure and to study the thermal stability,
annealing of selected samples deposited for \pn=20, 30, 50 and
100\% was carried out. Samples were annealed in a vacuum furnace
(pressure 1$\times$10$^{-6}$\,mabr) for about 1\,hour and XRD
measurements were carried out after each annealing as shown in
figure~\ref{fig:ann}(a)-(d), for \pn=20,30,50 and 100\%,
respectively. For \pn=20 and 30\%, XRD pattern are identical to
pristine state up to 423\,K. Above it, there are notable changes:
(i) the peak corresponding to (200) reflection of \tcn~shifts
towards higher angles (ii) a new peak corresponding to \tcn~(111)
starts to emerge. The peak shift upon annealing is more for
\pn=30\% sample and the intensity of (111) reflection is also more
prominent in this case. The peak shift towards higher angle is a
clear signature of contraction in $a$. The value of $a$ remains
constant at 0.370\,nm up to 423\,K and at 573\,K it becomes
0.356\,nm for \pn=30\p~sample. On the other hand, the sample
deposited at \pn=50\%, show very little changes with annealing
temperature, only peak width decreases, signifying grain growth
with annealing. For \pn=100\% sample, we find that the structure
is stable only up to 423\,K and at 473\,K, the broad peak observed
for the pristine sample, splits into two peaks, one corresponding
to CoN(200) and other to \tcn(111). Further annealing at 573\,K
results in formation of pure Co $hcp$ phase.

\begin{figure}\center
\includegraphics [width=85mm,height=70mm] {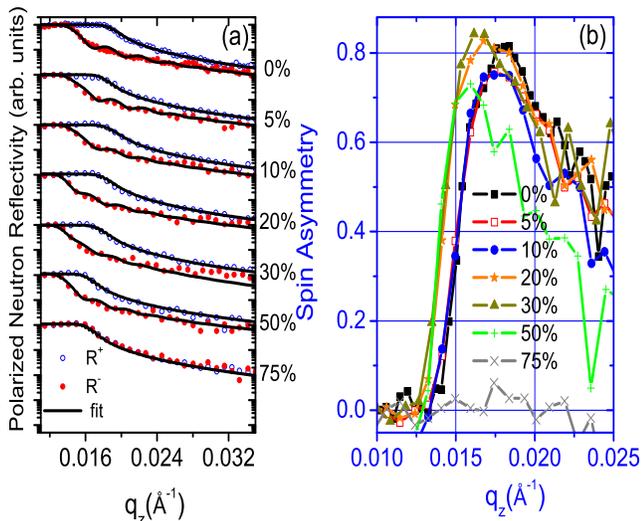} \vspace{-4mm}
\caption{\label{fig:pnr} (Color online) PNR pattern of Co-N thin
films deposited for various \pn (a) and spin asymmetry (b). PNR
patterns are shifted vertically, for clarity.}
\end{figure}

The XRD measurements carried out in the pristine and annealed
samples provide an insight about the phase formation the thermal
stability of samples. Broadly, observed behavior is in line with
those reported by Fang et al.~\cite{JVSTA:Fang:CoN} and Oda et
al.~\cite{JMS87_Oda} Interestingly, an incremental increase in the
$a$ of \tcn~for \pn=20 and 30\%, immediately opens up a
possibility to testify the large-volume high-moment approach for
\tcn. Under this model the relation between the $a$ and average
magnetic moment ($\mu$) is given by~\cite{JMMM10_Imai}
\begin{equation} \label{eq1}
a(x)=a_A(1-x)+a_Bx+C\mu,
\end{equation}

where $x$ is the atomic fraction, $a_A$, $a_B$ and $C$ are
parameters. For precise measurement of $\mu$, polarized neutron
reflectivity (PNR) is a well-known technique. Unlike bulk
magnetization techniques, it is independent of substrate
magnetization and sample mass.~\cite{PRB92_Blundell} It is
surprising to note that PNR has not yet been used for
quantification of $\mu$ in Co-N thin films. Neutrons being spin
$\frac{1}{2}$ particles have two states of quantization, parallel
(+) or antiparallel (-) to applied external field ($H$). Neutrons
interact with the magnetic field generated by unpaired spins on an
atomic magnet via dipolar interaction. The magnetic potential for
such kind of interaction is given by \cite{PRB97_Hope}
$V_m=-\mu_n\cdot\textbf{B}$. Where magnetic induction
$\textbf{B}=H+4\pi M$, $\mu_n$ is neutron magnetic moment and
$\textbf{M}$ is sample magnetization. We did PNR measurements at
Narziss neutron reflectometer at SINQ/PSI, Switzerland. During
measurements a magnetic field of about 0.2\,Tesla was applied
(parallel to sample surface) to saturate them magnetically.
Measured PNR for spin-up ($R^+$) and spin-down ($R^-$) can be
directly used for the two spin states of neutron using spin
asymmetry ($SA$), given by:~\cite{PRB92_Blundell} $SA =
(R^{+}-R^{-})/(R^{+}+R^{-})$. PNR pattern and $SA$ for samples
prepared using different \pn~are shown in fig.~\ref{fig:pnr} (a)
and (b), respectively.

Since $SA$ is a directly proportional to $\mu$, it can be compared
to see the relative changes as \pn~increases. Compared to \pn=0\%
sample, $SA$ decreases marginally for \pn=5 and 10\p~samples.
However for \pn=20 and 30\p~samples, $SA$ takes an upturn and its
maximum shifts towards lower $q_z$. While the shift to lower $q_z$
is due to overall reduction in film density, an increase in the
maximum value of $SA$, is a direct measure of enhanced $\mu$ as
compared to pure Co films. When \pn~is increased to 50\p, the
maxima in the $SA$ drops-off rapidly and for \pn=75\p, $R^{+} =
R^{-}$ and $SA\sim0$, indicating that the sample has become
non-ferromagnetic. For \pn=100\p, results (not shown) are similar
to \pn=75\p~case.

To get qualitative information of $\mu$, PNR pattern were fitted
using SimulReflec software~\cite{SimulReflec} and density of films
obtained from XRR measurements was taken as a input. Obtained
values of $\mu$ are given table~\ref{table1}. It can be seen as
\pn~is increases, first $\mu$ decreases and and than it increases
for \pn=20 and 30\p~samples. For \pn=50\%, it decreases again and
finally becomes zero for samples prepared at \pn=75 and 100\%.
From eq.\ref{eq1}, it is expected as $a$ increases, $\mu$ should
increase. Although such enhancement is theoretically
predicted~\cite{JAC14_Imai,PRB07_Matar,JMMM10_Matar}, it is
evidenced experimentally in this work for \tcn~thin films.

It may be noted that the experimentally obtained values of $a$ for
\tcn~are typically
0.357\,nm~\cite{JMS87_Oda,JAC14_Lourenco,JCS11_Ito}, much smaller
than the theoretically predicted value at
0.373\,nm~\cite{PRB07_Matar}. While for pure Co ($fcc$),
$a$=0.354. We find that for $a$=0.358\,nm(\pn=10\p~sample), the
value of $\mu$ is lower than that of pure Co, as observed in other
studies. However, for \pn=30\p, $a$=0.370\,nm and in this
condition, it is expected that $\mu$ of Co in \tcn~should be
larger than that of pure Co. Here it is interesting to note that
in most of the studies, \tcn~films have always been deposited at
elevated temperature between 433-723\,K. Theoretical calculation
predict that the heat of formation ($\Delta H_f^{\circ}$) for Co-N
system are even larger than than that of Fe-N
system~\cite{PRB83_Hf}). In this situation, when deposited at
elevated temperature, nitrogen deficient \tcn~films are obtained
characterized by smaller than theoretical values of $a$. When
deposited at room temperature, nitrogen incorporation within
\tcn~can be maximized leading to enhanced the magnetic moment as
observed in our samples.

In conclusion we studied the phase formation, thermal stability
and magnetization of Co-N thin films prepared using dcMS at room
temperature. We find that Co,\tcn,Co$_3$N and CoN phases are
formed as \pn~increases. While \tcn~and CoN phases are stable only
up to about 423\,K, Co$_3$N phase is more stable. More remarkable
results are obtained for \tcn~films. As \tcn~phase is formed, an
incremental increase in \pn~results in an expansion of lattice
constant, which in turn results in an enhanced Co magnetic moment
in \tcn~phase.


One of author (RG) would like thank UGC-DAE CSR, Indore for a CRS
project and to Ajay Gupta and V. Ganesan for support and
encouragements. A part of this work was performed at Narziss,
SINQ/PSI, Villigen, Switzerland. We are thankful to C. Schanzer
and M. Schneider for providing access to Narziss, M. Horisberger
for sputtering, V. R. Reddy and A. Gome for XRR and L. Behera for
help in XRD and SIMS measurements. We acknowledge DST, New Delhi
for providing financial support to carry out NR experiments.



\begin{thebibliography}{36}%
\makeatletter
\providecommand \@ifxundefined [1]{%
 \@ifx{#1\undefined}
}%
\providecommand \@ifnum [1]{%
 \ifnum #1\expandafter \@firstoftwo
 \else \expandafter \@secondoftwo
 \fi
}%
\providecommand \@ifx [1]{%
 \ifx #1\expandafter \@firstoftwo
 \else \expandafter \@secondoftwo
 \fi
}%
\providecommand \natexlab [1]{#1}%
\providecommand \enquote  [1]{``#1''}%
\providecommand \bibnamefont  [1]{#1}%
\providecommand \bibfnamefont [1]{#1}%
\providecommand \citenamefont [1]{#1}%
\providecommand \href@noop [0]{\@secondoftwo}%
\providecommand \href [0]{\begingroup \@sanitize@url \@href}%
\providecommand \@href[1]{\@@startlink{#1}\@@href}%
\providecommand \@@href[1]{\endgroup#1\@@endlink}%
\providecommand \@sanitize@url [0]{\catcode `\\12\catcode
`\$12\catcode
  `\&12\catcode `\#12\catcode `\^12\catcode `\_12\catcode `\%12\relax}%
\providecommand \@@startlink[1]{}%
\providecommand \@@endlink[0]{}%
\providecommand \url  [0]{\begingroup\@sanitize@url \@url }%
\providecommand \@url [1]{\endgroup\@href {#1}{\urlprefix }}%
\providecommand \urlprefix  [0]{URL }%
\providecommand \Eprint [0]{\href }%
\providecommand \doibase [0]{http://dx.doi.org/}%
\providecommand \selectlanguage [0]{\@gobble}%
\providecommand \bibinfo  [0]{\@secondoftwo}%
\providecommand \bibfield  [0]{\@secondoftwo}%
\providecommand \translation [1]{[#1]}%
\providecommand \BibitemOpen [0]{}%
\providecommand \bibitemStop [0]{}%
\providecommand \bibitemNoStop [0]{.\EOS\space}%
\providecommand \EOS [0]{\spacefactor3000\relax}%
\providecommand \BibitemShut  [1]{\csname bibitem#1\endcsname}%
\let\auto@bib@innerbib\@empty
\bibitem [{\citenamefont {Ji}\ \emph {et~al.}(2011)\citenamefont {Ji},
  \citenamefont {Allard}, \citenamefont {Lara-Curzio},\ and\ \citenamefont
  {Wang}}]{APL11_Ji}%
  \BibitemOpen
  \bibfield  {author} {\bibinfo {author} {\bibfnamefont {N.}~\bibnamefont
  {Ji}}, \bibinfo {author} {\bibfnamefont {L.~F.}\ \bibnamefont {Allard}},
  \bibinfo {author} {\bibfnamefont {E.}~\bibnamefont {Lara-Curzio}}, \ and\
  \bibinfo {author} {\bibfnamefont {J.-P.}\ \bibnamefont {Wang}},\ }\href
  {\doibase http://dx.doi.org/10.1063/1.3560051} {\bibfield  {journal}
  {\bibinfo  {journal} {Applied Physics Letters}\ }\textbf {\bibinfo {volume}
  {98}},\ \bibinfo {eid} {092506} (\bibinfo {year} {2011})}\BibitemShut
  {NoStop}%
\bibitem [{\citenamefont {Ito}\ \emph {et~al.}(2011{\natexlab{a}})\citenamefont
  {Ito}, \citenamefont {Lee}, \citenamefont {Harada}, \citenamefont {Suzuno},
  \citenamefont {Suemasu}, \citenamefont {Takeda}, \citenamefont {Saitoh},
  \citenamefont {Ye}, \citenamefont {Kimura},\ and\ \citenamefont
  {Akinaga}}]{APL11_Ito_Fe4N}%
  \BibitemOpen
  \bibfield  {author} {\bibinfo {author} {\bibfnamefont {K.}~\bibnamefont
  {Ito}}, \bibinfo {author} {\bibfnamefont {G.~H.}\ \bibnamefont {Lee}},
  \bibinfo {author} {\bibfnamefont {K.}~\bibnamefont {Harada}}, \bibinfo
  {author} {\bibfnamefont {M.}~\bibnamefont {Suzuno}}, \bibinfo {author}
  {\bibfnamefont {T.}~\bibnamefont {Suemasu}}, \bibinfo {author} {\bibfnamefont
  {Y.}~\bibnamefont {Takeda}}, \bibinfo {author} {\bibfnamefont
  {Y.}~\bibnamefont {Saitoh}}, \bibinfo {author} {\bibfnamefont
  {M.}~\bibnamefont {Ye}}, \bibinfo {author} {\bibfnamefont {A.}~\bibnamefont
  {Kimura}}, \ and\ \bibinfo {author} {\bibfnamefont {H.}~\bibnamefont
  {Akinaga}},\ }\href {\doibase http://dx.doi.org/10.1063/1.3564887} {\bibfield
   {journal} {\bibinfo  {journal} {Applied Physics Letters}\ }\textbf {\bibinfo
  {volume} {98}},\ \bibinfo {eid} {102507} (\bibinfo {year}
  {2011}{\natexlab{a}})}\BibitemShut {NoStop}%
\bibitem [{\citenamefont {Atiq}\ \emph {et~al.}(2008)\citenamefont {Atiq},
  \citenamefont {Ko}, \citenamefont {Siddiqi},\ and\ \citenamefont
  {Shin}}]{APL08_Atiq}%
  \BibitemOpen
  \bibfield  {author} {\bibinfo {author} {\bibfnamefont {S.}~\bibnamefont
  {Atiq}}, \bibinfo {author} {\bibfnamefont {H.-S.}\ \bibnamefont {Ko}},
  \bibinfo {author} {\bibfnamefont {S.~A.}\ \bibnamefont {Siddiqi}}, \ and\
  \bibinfo {author} {\bibfnamefont {S.-C.}\ \bibnamefont {Shin}},\ }\href
  {\doibase http://dx.doi.org/10.1063/1.2940599} {\bibfield  {journal}
  {\bibinfo  {journal} {Applied Physics Letters}\ }\textbf {\bibinfo {volume}
  {92}},\ \bibinfo {eid} {222507} (\bibinfo {year} {2008})}\BibitemShut
  {NoStop}%
\bibitem [{\citenamefont {Bhattacharyya}(2015)}]{JPCC15_Bhatta}%
  \BibitemOpen
  \bibfield  {author} {\bibinfo {author} {\bibfnamefont {S.}~\bibnamefont
  {Bhattacharyya}},\ }\href {\doibase 10.1021/jp510606z} {\bibfield  {journal}
  {\bibinfo  {journal} {The Journal of Physical Chemistry C}\ }\textbf
  {\bibinfo {volume} {119}},\ \bibinfo {pages} {1601} (\bibinfo {year}
  {2015})}\BibitemShut {NoStop}%
\bibitem [{\citenamefont {Tayal}\ \emph {et~al.}(2014)\citenamefont {Tayal},
  \citenamefont {Gupta}, \citenamefont {Lalla}, \citenamefont {Gupta},
  \citenamefont {Horisberger}, \citenamefont {Stahn}, \citenamefont {Schlage},\
  and\ \citenamefont {Wille}}]{PRB14_MG}%
  \BibitemOpen
  \bibfield  {author} {\bibinfo {author} {\bibfnamefont {A.}~\bibnamefont
  {Tayal}}, \bibinfo {author} {\bibfnamefont {M.}~\bibnamefont {Gupta}},
  \bibinfo {author} {\bibfnamefont {N.~P.}\ \bibnamefont {Lalla}}, \bibinfo
  {author} {\bibfnamefont {A.}~\bibnamefont {Gupta}}, \bibinfo {author}
  {\bibfnamefont {M.}~\bibnamefont {Horisberger}}, \bibinfo {author}
  {\bibfnamefont {J.}~\bibnamefont {Stahn}}, \bibinfo {author} {\bibfnamefont
  {K.}~\bibnamefont {Schlage}}, \ and\ \bibinfo {author} {\bibfnamefont
  {H.-C.}\ \bibnamefont {Wille}},\ }\href {\doibase 10.1103/PhysRevB.90.144412}
  {\bibfield  {journal} {\bibinfo  {journal} {Phys. Rev. B}\ }\textbf {\bibinfo
  {volume} {90}},\ \bibinfo {pages} {144412} (\bibinfo {year}
  {2014})}\BibitemShut {NoStop}%
\bibitem [{\citenamefont {Imai}, \citenamefont {Sohma},\ and\ \citenamefont
  {Suemasu}(2014)}]{JAC14_Imai}%
  \BibitemOpen
  \bibfield  {author} {\bibinfo {author} {\bibfnamefont {Y.}~\bibnamefont
  {Imai}}, \bibinfo {author} {\bibfnamefont {M.}~\bibnamefont {Sohma}}, \ and\
  \bibinfo {author} {\bibfnamefont {T.}~\bibnamefont {Suemasu}},\ }\href
  {\doibase http://dx.doi.org/10.1016/j.jallcom.2014.04.171} {\bibfield
  {journal} {\bibinfo  {journal} {Journal of Alloys and Compounds}\ }\textbf
  {\bibinfo {volume} {611}},\ \bibinfo {pages} {440 } (\bibinfo {year}
  {2014})}\BibitemShut {NoStop}%
\bibitem [{\citenamefont {Matar}, \citenamefont {Houari},\ and\ \citenamefont
  {Belkhir}(2007)}]{PRB07_Matar}%
  \BibitemOpen
  \bibfield  {author} {\bibinfo {author} {\bibfnamefont {S.~F.}\ \bibnamefont
  {Matar}}, \bibinfo {author} {\bibfnamefont {A.}~\bibnamefont {Houari}}, \
  and\ \bibinfo {author} {\bibfnamefont {M.~A.}\ \bibnamefont {Belkhir}},\
  }\href {\doibase 10.1103/PhysRevB.75.245109} {\bibfield  {journal} {\bibinfo
  {journal} {Phys. Rev. B}\ }\textbf {\bibinfo {volume} {75}},\ \bibinfo
  {pages} {245109} (\bibinfo {year} {2007})}\BibitemShut {NoStop}%
\bibitem [{\citenamefont {Matsuoka}, \citenamefont {Ono},\ and\ \citenamefont
  {Inukai}(1986)}]{APL86_Matsuoka}%
  \BibitemOpen
  \bibfield  {author} {\bibinfo {author} {\bibfnamefont {M.}~\bibnamefont
  {Matsuoka}}, \bibinfo {author} {\bibfnamefont {K.}~\bibnamefont {Ono}}, \
  and\ \bibinfo {author} {\bibfnamefont {T.}~\bibnamefont {Inukai}},\
  }\href@noop {} {\bibfield  {journal} {\bibinfo  {journal} {Applied Physics
  Letters}\ }\textbf {\bibinfo {volume} {49}} (\bibinfo {year}
  {1986})}\BibitemShut {NoStop}%
\bibitem [{\citenamefont {Oda}, \citenamefont {Yoshio},\ and\ \citenamefont
  {Oda}(1987)}]{JMS87_Oda}%
  \BibitemOpen
  \bibfield  {author} {\bibinfo {author} {\bibfnamefont {K.}~\bibnamefont
  {Oda}}, \bibinfo {author} {\bibfnamefont {T.}~\bibnamefont {Yoshio}}, \ and\
  \bibinfo {author} {\bibfnamefont {K.}~\bibnamefont {Oda}},\ }\href {\doibase
  10.1007/BF01086464} {\bibfield  {journal} {\bibinfo  {journal} {Journal of
  Materials Science}\ }\textbf {\bibinfo {volume} {22}},\ \bibinfo {pages}
  {2729} (\bibinfo {year} {1987})}\BibitemShut {NoStop}%
\bibitem [{\citenamefont {Suzuki}\ \emph {et~al.}(1995)\citenamefont {Suzuki},
  \citenamefont {Kaneko}, \citenamefont {Yoshida}, \citenamefont {Morita},\
  and\ \citenamefont {Fujimori}}]{JAC95_Suzuki}%
  \BibitemOpen
  \bibfield  {author} {\bibinfo {author} {\bibfnamefont {K.}~\bibnamefont
  {Suzuki}}, \bibinfo {author} {\bibfnamefont {T.}~\bibnamefont {Kaneko}},
  \bibinfo {author} {\bibfnamefont {H.}~\bibnamefont {Yoshida}}, \bibinfo
  {author} {\bibfnamefont {H.}~\bibnamefont {Morita}}, \ and\ \bibinfo {author}
  {\bibfnamefont {H.}~\bibnamefont {Fujimori}},\ }\href {\doibase
  http://dx.doi.org/10.1016/0925-8388(95)01561-2} {\bibfield  {journal}
  {\bibinfo  {journal} {Journal of Alloys and Compounds}\ }\textbf {\bibinfo
  {volume} {224}},\ \bibinfo {pages} {232 } (\bibinfo {year}
  {1995})}\BibitemShut {NoStop}%
\bibitem [{\citenamefont {Asahara}\ \emph {et~al.}(2001)\citenamefont
  {Asahara}, \citenamefont {Migita}, \citenamefont {Tanaka},\ and\
  \citenamefont {Kawabata}}]{Vac01_Asahara}%
  \BibitemOpen
  \bibfield  {author} {\bibinfo {author} {\bibfnamefont {H.}~\bibnamefont
  {Asahara}}, \bibinfo {author} {\bibfnamefont {T.}~\bibnamefont {Migita}},
  \bibinfo {author} {\bibfnamefont {T.}~\bibnamefont {Tanaka}}, \ and\ \bibinfo
  {author} {\bibfnamefont {K.}~\bibnamefont {Kawabata}},\ }\href {\doibase
  http://dx.doi.org/10.1016/S0042-207X(00)00453-X} {\bibfield  {journal}
  {\bibinfo  {journal} {Vacuum}\ }\textbf {\bibinfo {volume} {62}},\ \bibinfo
  {pages} {293 } (\bibinfo {year} {2001})}\BibitemShut {NoStop}%
\bibitem [{\citenamefont {Fang}\ \emph {et~al.}(2004)\citenamefont {Fang},
  \citenamefont {Yang}, \citenamefont {Hsu}, \citenamefont {Chen},
  \citenamefont {Lin},\ and\ \citenamefont {Chen}}]{JVSTA:Fang:CoN}%
  \BibitemOpen
  \bibfield  {author} {\bibinfo {author} {\bibfnamefont {J.-S.}\ \bibnamefont
  {Fang}}, \bibinfo {author} {\bibfnamefont {L.-C.}\ \bibnamefont {Yang}},
  \bibinfo {author} {\bibfnamefont {C.-S.}\ \bibnamefont {Hsu}}, \bibinfo
  {author} {\bibfnamefont {G.-S.}\ \bibnamefont {Chen}}, \bibinfo {author}
  {\bibfnamefont {Y.-W.}\ \bibnamefont {Lin}}, \ and\ \bibinfo {author}
  {\bibfnamefont {G.-S.}\ \bibnamefont {Chen}},\ }\href@noop {} {\bibfield
  {journal} {\bibinfo  {journal} {Journal of Vacuum Science \& Technology A}\
  }\textbf {\bibinfo {volume} {22}} (\bibinfo {year} {2004})}\BibitemShut
  {NoStop}%
\bibitem [{\citenamefont {Jia}\ \emph {et~al.}(2008)\citenamefont {Jia},
  \citenamefont {Wang}, \citenamefont {Zheng}, \citenamefont {Chen},\ and\
  \citenamefont {Feng}}]{MSEB08_Jia}%
  \BibitemOpen
  \bibfield  {author} {\bibinfo {author} {\bibfnamefont {H.}~\bibnamefont
  {Jia}}, \bibinfo {author} {\bibfnamefont {X.}~\bibnamefont {Wang}}, \bibinfo
  {author} {\bibfnamefont {W.}~\bibnamefont {Zheng}}, \bibinfo {author}
  {\bibfnamefont {Y.}~\bibnamefont {Chen}}, \ and\ \bibinfo {author}
  {\bibfnamefont {S.}~\bibnamefont {Feng}},\ }\href {\doibase
  http://dx.doi.org/10.1016/j.mseb.2008.03.001} {\bibfield  {journal} {\bibinfo
   {journal} {Materials Science and Engineering: B}\ }\textbf {\bibinfo
  {volume} {150}},\ \bibinfo {pages} {121 } (\bibinfo {year}
  {2008})}\BibitemShut {NoStop}%
\bibitem [{\citenamefont {Paduani}(2008)}]{SSC08_Paduani}%
  \BibitemOpen
  \bibfield  {author} {\bibinfo {author} {\bibfnamefont {C.}~\bibnamefont
  {Paduani}},\ }\href {\doibase http://dx.doi.org/10.1016/j.ssc.2008.09.010}
  {\bibfield  {journal} {\bibinfo  {journal} {Solid State Communications}\
  }\textbf {\bibinfo {volume} {148}},\ \bibinfo {pages} {297 } (\bibinfo {year}
  {2008})}\BibitemShut {NoStop}%
\bibitem [{\citenamefont {Houari}, \citenamefont {Matar},\ and\ \citenamefont
  {Belkhir}(2010)}]{JMMM10_Matar}%
  \BibitemOpen
  \bibfield  {author} {\bibinfo {author} {\bibfnamefont {A.}~\bibnamefont
  {Houari}}, \bibinfo {author} {\bibfnamefont {S.~F.}\ \bibnamefont {Matar}}, \
  and\ \bibinfo {author} {\bibfnamefont {M.~A.}\ \bibnamefont {Belkhir}},\
  }\href {\doibase http://dx.doi.org/10.1016/j.jmmm.2009.10.034} {\bibfield
  {journal} {\bibinfo  {journal} {Journal of Magnetism and Magnetic Materials}\
  }\textbf {\bibinfo {volume} {322}},\ \bibinfo {pages} {658 } (\bibinfo {year}
  {2010})}\BibitemShut {NoStop}%
\bibitem [{\citenamefont {Imai}, \citenamefont {Takahashi},\ and\ \citenamefont
  {Kumagai}(2010)}]{JMMM10_Imai}%
  \BibitemOpen
  \bibfield  {author} {\bibinfo {author} {\bibfnamefont {Y.}~\bibnamefont
  {Imai}}, \bibinfo {author} {\bibfnamefont {Y.}~\bibnamefont {Takahashi}}, \
  and\ \bibinfo {author} {\bibfnamefont {T.}~\bibnamefont {Kumagai}},\ }\href
  {\doibase http://dx.doi.org/10.1016/j.jmmm.2010.04.004} {\bibfield  {journal}
  {\bibinfo  {journal} {Journal of Magnetism and Magnetic Materials}\ }\textbf
  {\bibinfo {volume} {322}},\ \bibinfo {pages} {2665 } (\bibinfo {year}
  {2010})}\BibitemShut {NoStop}%
\bibitem [{\citenamefont {Takahashi}, \citenamefont {Imai},\ and\ \citenamefont
  {Kumagai}(2011)}]{JMMM11_Takahashi}%
  \BibitemOpen
  \bibfield  {author} {\bibinfo {author} {\bibfnamefont {Y.}~\bibnamefont
  {Takahashi}}, \bibinfo {author} {\bibfnamefont {Y.}~\bibnamefont {Imai}}, \
  and\ \bibinfo {author} {\bibfnamefont {T.}~\bibnamefont {Kumagai}},\ }\href
  {\doibase http://dx.doi.org/10.1016/j.jmmm.2011.05.030} {\bibfield  {journal}
  {\bibinfo  {journal} {Journal of Magnetism and Magnetic Materials}\ }\textbf
  {\bibinfo {volume} {323}},\ \bibinfo {pages} {2941 } (\bibinfo {year}
  {2011})}\BibitemShut {NoStop}%
\bibitem [{\citenamefont {Ito}\ \emph {et~al.}(2011{\natexlab{b}})\citenamefont
  {Ito}, \citenamefont {Harada}, \citenamefont {Toko}, \citenamefont {Ye},
  \citenamefont {Kimura}, \citenamefont {Takeda}, \citenamefont {Saitoh},
  \citenamefont {Akinaga},\ and\ \citenamefont {Suemasu}}]{APL11_Ito_Co4N}%
  \BibitemOpen
  \bibfield  {author} {\bibinfo {author} {\bibfnamefont {K.}~\bibnamefont
  {Ito}}, \bibinfo {author} {\bibfnamefont {K.}~\bibnamefont {Harada}},
  \bibinfo {author} {\bibfnamefont {K.}~\bibnamefont {Toko}}, \bibinfo {author}
  {\bibfnamefont {M.}~\bibnamefont {Ye}}, \bibinfo {author} {\bibfnamefont
  {A.}~\bibnamefont {Kimura}}, \bibinfo {author} {\bibfnamefont
  {Y.}~\bibnamefont {Takeda}}, \bibinfo {author} {\bibfnamefont
  {Y.}~\bibnamefont {Saitoh}}, \bibinfo {author} {\bibfnamefont
  {H.}~\bibnamefont {Akinaga}}, \ and\ \bibinfo {author} {\bibfnamefont
  {T.}~\bibnamefont {Suemasu}},\ }\href {\doibase
  http://dx.doi.org/10.1063/1.3670353} {\bibfield  {journal} {\bibinfo
  {journal} {Applied Physics Letters}\ }\textbf {\bibinfo {volume} {99}},\
  \bibinfo {eid} {252501} (\bibinfo {year} {2011}{\natexlab{b}})}\BibitemShut
  {NoStop}%
\bibitem [{\citenamefont {Ito}\ \emph {et~al.}(2011{\natexlab{c}})\citenamefont
  {Ito}, \citenamefont {Harada}, \citenamefont {Toko}, \citenamefont
  {Akinaga},\ and\ \citenamefont {Suemasu}}]{JCS11_Ito}%
  \BibitemOpen
  \bibfield  {author} {\bibinfo {author} {\bibfnamefont {K.}~\bibnamefont
  {Ito}}, \bibinfo {author} {\bibfnamefont {K.}~\bibnamefont {Harada}},
  \bibinfo {author} {\bibfnamefont {K.}~\bibnamefont {Toko}}, \bibinfo {author}
  {\bibfnamefont {H.}~\bibnamefont {Akinaga}}, \ and\ \bibinfo {author}
  {\bibfnamefont {T.}~\bibnamefont {Suemasu}},\ }\href {\doibase
  http://dx.doi.org/10.1016/j.jcrysgro.2011.09.038} {\bibfield  {journal}
  {\bibinfo  {journal} {Journal of Crystal Growth}\ }\textbf {\bibinfo {volume}
  {336}},\ \bibinfo {pages} {40 } (\bibinfo {year}
  {2011}{\natexlab{c}})}\BibitemShut {NoStop}%
\bibitem [{\citenamefont {Bhandari}\ \emph {et~al.}(2012)\citenamefont
  {Bhandari}, \citenamefont {Yang}, \citenamefont {Kim}, \citenamefont {Lin},
  \citenamefont {Gordon}, \citenamefont {Wang}, \citenamefont {Lehn},
  \citenamefont {Li},\ and\ \citenamefont {Shenai}}]{ECS12_Bhandari}%
  \BibitemOpen
  \bibfield  {author} {\bibinfo {author} {\bibfnamefont {H.~B.}\ \bibnamefont
  {Bhandari}}, \bibinfo {author} {\bibfnamefont {J.}~\bibnamefont {Yang}},
  \bibinfo {author} {\bibfnamefont {H.}~\bibnamefont {Kim}}, \bibinfo {author}
  {\bibfnamefont {Y.}~\bibnamefont {Lin}}, \bibinfo {author} {\bibfnamefont
  {R.~G.}\ \bibnamefont {Gordon}}, \bibinfo {author} {\bibfnamefont {Q.~M.}\
  \bibnamefont {Wang}}, \bibinfo {author} {\bibfnamefont {J.-S.~M.}\
  \bibnamefont {Lehn}}, \bibinfo {author} {\bibfnamefont {H.}~\bibnamefont
  {Li}}, \ and\ \bibinfo {author} {\bibfnamefont {D.}~\bibnamefont {Shenai}},\
  }\href {\doibase 10.1149/2.005205jss} {\bibfield  {journal} {\bibinfo
  {journal} {ECS Journal of Solid State Science and Technology}\ }\textbf
  {\bibinfo {volume} {1}},\ \bibinfo {pages} {N79} (\bibinfo {year}
  {2012})}\BibitemShut {NoStop}%
\bibitem [{\citenamefont {Louren\c{c}o}\ \emph {et~al.}(2014)\citenamefont
  {Louren\c{c}o}, \citenamefont {Carvalho}, \citenamefont {Fonseca},
  \citenamefont {Gasche}, \citenamefont {Evans}, \citenamefont {Godinho},\ and\
  \citenamefont {Cruz}}]{JAC14_Lourenco}%
  \BibitemOpen
  \bibfield  {author} {\bibinfo {author} {\bibfnamefont {M.}~\bibnamefont
  {Louren\c{c}o}}, \bibinfo {author} {\bibfnamefont {M.}~\bibnamefont
  {Carvalho}}, \bibinfo {author} {\bibfnamefont {P.}~\bibnamefont {Fonseca}},
  \bibinfo {author} {\bibfnamefont {T.}~\bibnamefont {Gasche}}, \bibinfo
  {author} {\bibfnamefont {G.}~\bibnamefont {Evans}}, \bibinfo {author}
  {\bibfnamefont {M.}~\bibnamefont {Godinho}}, \ and\ \bibinfo {author}
  {\bibfnamefont {M.}~\bibnamefont {Cruz}},\ }\href {\doibase
  http://dx.doi.org/10.1016/j.jallcom.2014.05.048} {\bibfield  {journal}
  {\bibinfo  {journal} {Journal of Alloys and Compounds}\ }\textbf {\bibinfo
  {volume} {612}},\ \bibinfo {pages} {176 } (\bibinfo {year}
  {2014})}\BibitemShut {NoStop}%
\bibitem [{\citenamefont {Liu}\ \emph {et~al.}(2014)\citenamefont {Liu},
  \citenamefont {Lu}, \citenamefont {He}, \citenamefont {Jin}, \citenamefont
  {Yang}, \citenamefont {Ni},\ and\ \citenamefont {Zhao}}]{JAC14_Liu}%
  \BibitemOpen
  \bibfield  {author} {\bibinfo {author} {\bibfnamefont {X.}~\bibnamefont
  {Liu}}, \bibinfo {author} {\bibfnamefont {H.}~\bibnamefont {Lu}}, \bibinfo
  {author} {\bibfnamefont {M.}~\bibnamefont {He}}, \bibinfo {author}
  {\bibfnamefont {K.}~\bibnamefont {Jin}}, \bibinfo {author} {\bibfnamefont
  {G.}~\bibnamefont {Yang}}, \bibinfo {author} {\bibfnamefont {H.}~\bibnamefont
  {Ni}}, \ and\ \bibinfo {author} {\bibfnamefont {K.}~\bibnamefont {Zhao}},\
  }\href {\doibase http://dx.doi.org/10.1016/j.jallcom.2013.08.001} {\bibfield
  {journal} {\bibinfo  {journal} {Journal of Alloys and Compounds}\ }\textbf
  {\bibinfo {volume} {582}},\ \bibinfo {pages} {75 } (\bibinfo {year}
  {2014})}\BibitemShut {NoStop}%
\bibitem [{\citenamefont {Silva}\ \emph {et~al.}(2014)\citenamefont {Silva},
  \citenamefont {Vovk}, \citenamefont {da~Silva}, \citenamefont {Strichovanec},
  \citenamefont {Algarabel}, \citenamefont {Gonçalves}, \citenamefont
  {Borges}, \citenamefont {Godinho},\ and\ \citenamefont {Cruz}}]{TSF14_Silva}%
  \BibitemOpen
  \bibfield  {author} {\bibinfo {author} {\bibfnamefont {C.}~\bibnamefont
  {Silva}}, \bibinfo {author} {\bibfnamefont {A.}~\bibnamefont {Vovk}},
  \bibinfo {author} {\bibfnamefont {R.}~\bibnamefont {da~Silva}}, \bibinfo
  {author} {\bibfnamefont {P.}~\bibnamefont {Strichovanec}}, \bibinfo {author}
  {\bibfnamefont {P.}~\bibnamefont {Algarabel}}, \bibinfo {author}
  {\bibfnamefont {A.}~\bibnamefont {Gonçalves}}, \bibinfo {author}
  {\bibfnamefont {R.}~\bibnamefont {Borges}}, \bibinfo {author} {\bibfnamefont
  {M.}~\bibnamefont {Godinho}}, \ and\ \bibinfo {author} {\bibfnamefont
  {M.}~\bibnamefont {Cruz}},\ }\href {\doibase
  http://dx.doi.org/10.1016/j.tsf.2014.01.019} {\bibfield  {journal} {\bibinfo
  {journal} {Thin Solid Films}\ }\textbf {\bibinfo {volume} {556}},\ \bibinfo
  {pages} {125 } (\bibinfo {year} {2014})}\BibitemShut {NoStop}%
\bibitem [{\citenamefont {Silva}\ \emph {et~al.}(2015)\citenamefont {Silva},
  \citenamefont {Vovk}, \citenamefont {da~Silva}, \citenamefont {Strichonavec},
  \citenamefont {Algarabel}, \citenamefont {Casaca}, \citenamefont {Meneghini},
  \citenamefont {Carlomagno}, \citenamefont {Godinho},\ and\ \citenamefont
  {Cruz}}]{JAC15_Silva}%
  \BibitemOpen
  \bibfield  {author} {\bibinfo {author} {\bibfnamefont {C.}~\bibnamefont
  {Silva}}, \bibinfo {author} {\bibfnamefont {A.}~\bibnamefont {Vovk}},
  \bibinfo {author} {\bibfnamefont {R.}~\bibnamefont {da~Silva}}, \bibinfo
  {author} {\bibfnamefont {P.}~\bibnamefont {Strichonavec}}, \bibinfo {author}
  {\bibfnamefont {P.}~\bibnamefont {Algarabel}}, \bibinfo {author}
  {\bibfnamefont {A.}~\bibnamefont {Casaca}}, \bibinfo {author} {\bibfnamefont
  {C.}~\bibnamefont {Meneghini}}, \bibinfo {author} {\bibfnamefont
  {I.}~\bibnamefont {Carlomagno}}, \bibinfo {author} {\bibfnamefont
  {M.}~\bibnamefont {Godinho}}, \ and\ \bibinfo {author} {\bibfnamefont
  {M.}~\bibnamefont {Cruz}},\ }\href {\doibase
  http://dx.doi.org/10.1016/j.jallcom.2015.02.013} {\bibfield  {journal}
  {\bibinfo  {journal} {Journal of Alloys and Compounds}\ }\textbf {\bibinfo
  {volume} {633}},\ \bibinfo {pages} {470 } (\bibinfo {year}
  {2015})}\BibitemShut {NoStop}%
\bibitem [{\citenamefont {Braun}(7 99)}]{Parratt32}%
  \BibitemOpen
  \bibfield  {author} {\bibinfo {author} {\bibfnamefont {C.}~\bibnamefont
  {Braun}},\ }\href@noop {} {\emph {\bibinfo {title} {Parratt32- The
  Reflectivity Tool}}}\ (\bibinfo  {publisher} {HMI Berlin},\ \bibinfo {year}
  {1997-99})\BibitemShut {NoStop}%
\bibitem [{\citenamefont {Zhang}\ \emph {et~al.}(1999)\citenamefont {Zhang},
  \citenamefont {Poole}, \citenamefont {Eller},\ and\ \citenamefont
  {Keefe}}]{JVST99_Zhang}%
  \BibitemOpen
  \bibfield  {author} {\bibinfo {author} {\bibfnamefont {H.}~\bibnamefont
  {Zhang}}, \bibinfo {author} {\bibfnamefont {J.}~\bibnamefont {Poole}},
  \bibinfo {author} {\bibfnamefont {R.}~\bibnamefont {Eller}}, \ and\ \bibinfo
  {author} {\bibfnamefont {M.}~\bibnamefont {Keefe}},\ }\href@noop {}
  {\bibfield  {journal} {\bibinfo  {journal} {Journal of Vacuum Science \&
  Technology A}\ }\textbf {\bibinfo {volume} {17}} (\bibinfo {year}
  {1999})}\BibitemShut {NoStop}%
\bibitem [{\citenamefont {Yoo}\ \emph {et~al.}(2000)\citenamefont {Yoo},
  \citenamefont {Cynn}, \citenamefont {S\"oderlind},\ and\ \citenamefont
  {Iota}}]{PRL00_newCo}%
  \BibitemOpen
  \bibfield  {author} {\bibinfo {author} {\bibfnamefont {C.~S.}\ \bibnamefont
  {Yoo}}, \bibinfo {author} {\bibfnamefont {H.}~\bibnamefont {Cynn}}, \bibinfo
  {author} {\bibfnamefont {P.}~\bibnamefont {S\"oderlind}}, \ and\ \bibinfo
  {author} {\bibfnamefont {V.}~\bibnamefont {Iota}},\ }\href {\doibase
  10.1103/PhysRevLett.84.4132} {\bibfield  {journal} {\bibinfo  {journal}
  {Phys. Rev. Lett.}\ }\textbf {\bibinfo {volume} {84}},\ \bibinfo {pages}
  {4132} (\bibinfo {year} {2000})}\BibitemShut {NoStop}%
\bibitem [{\citenamefont {Ko}\ \emph {et~al.}(2003)\citenamefont {Ko},
  \citenamefont {Park}, \citenamefont {Seo}, \citenamefont {Yang},
  \citenamefont {Shin}, \citenamefont {Kim}, \citenamefont {Lee}, \citenamefont
  {Lee}, \citenamefont {Reucroft},\ and\ \citenamefont {Lee}}]{MCP03_Ko}%
  \BibitemOpen
  \bibfield  {author} {\bibinfo {author} {\bibfnamefont {Y.}~\bibnamefont
  {Ko}}, \bibinfo {author} {\bibfnamefont {D.}~\bibnamefont {Park}}, \bibinfo
  {author} {\bibfnamefont {B.}~\bibnamefont {Seo}}, \bibinfo {author}
  {\bibfnamefont {H.}~\bibnamefont {Yang}}, \bibinfo {author} {\bibfnamefont
  {H.}~\bibnamefont {Shin}}, \bibinfo {author} {\bibfnamefont {J.}~\bibnamefont
  {Kim}}, \bibinfo {author} {\bibfnamefont {J.}~\bibnamefont {Lee}}, \bibinfo
  {author} {\bibfnamefont {W.}~\bibnamefont {Lee}}, \bibinfo {author}
  {\bibfnamefont {P.}~\bibnamefont {Reucroft}}, \ and\ \bibinfo {author}
  {\bibfnamefont {J.}~\bibnamefont {Lee}},\ }\href {\doibase
  http://dx.doi.org/10.1016/S0254-0584(03)00085-3} {\bibfield  {journal}
  {\bibinfo  {journal} {Materials Chemistry and Physics}\ }\textbf {\bibinfo
  {volume} {80}},\ \bibinfo {pages} {560 } (\bibinfo {year}
  {2003})}\BibitemShut {NoStop}%
\bibitem [{\citenamefont {Vaz}, \citenamefont {Bland},\ and\ \citenamefont
  {Lauhoff}(2008)}]{RPP08_Vaz}%
  \BibitemOpen
  \bibfield  {author} {\bibinfo {author} {\bibfnamefont {C.~A.~F.}\
  \bibnamefont {Vaz}}, \bibinfo {author} {\bibfnamefont {J.~A.~C.}\
  \bibnamefont {Bland}}, \ and\ \bibinfo {author} {\bibfnamefont
  {G.}~\bibnamefont {Lauhoff}},\ }\href
  {http://stacks.iop.org/0034-4885/71/i=5/a=056501} {\bibfield  {journal}
  {\bibinfo  {journal} {Reports on Progress in Physics}\ }\textbf {\bibinfo
  {volume} {71}},\ \bibinfo {pages} {056501} (\bibinfo {year}
  {2008})}\BibitemShut {NoStop}%
\bibitem [{\citenamefont {Gupta}\ and\ \citenamefont
  {Gupta}(2005)}]{Gupta:PRB05}%
  \BibitemOpen
  \bibfield  {author} {\bibinfo {author} {\bibfnamefont {R.}~\bibnamefont
  {Gupta}}\ and\ \bibinfo {author} {\bibfnamefont {M.}~\bibnamefont {Gupta}},\
  }\href {\doibase doi:10.1103/PhysRevB.72.024202} {\bibfield  {journal}
  {\bibinfo  {journal} {Phys. Rev. B}\ }\textbf {\bibinfo {volume} {72}},\
  \bibinfo {pages} {024202} (\bibinfo {year} {2005})}\BibitemShut {NoStop}%
\bibitem [{\citenamefont {Gupta}\ \emph
  {et~al.}(2011{\natexlab{a}})\citenamefont {Gupta}, \citenamefont {Tayal},
  \citenamefont {Gupta}, \citenamefont {Reddy}, \citenamefont {Horisberger},\
  and\ \citenamefont {Stahn}}]{MG:JAC:2011}%
  \BibitemOpen
  \bibfield  {author} {\bibinfo {author} {\bibfnamefont {M.}~\bibnamefont
  {Gupta}}, \bibinfo {author} {\bibfnamefont {A.}~\bibnamefont {Tayal}},
  \bibinfo {author} {\bibfnamefont {A.}~\bibnamefont {Gupta}}, \bibinfo
  {author} {\bibfnamefont {V.}~\bibnamefont {Reddy}}, \bibinfo {author}
  {\bibfnamefont {M.}~\bibnamefont {Horisberger}}, \ and\ \bibinfo {author}
  {\bibfnamefont {J.}~\bibnamefont {Stahn}},\ }\href {\doibase
  10.1016/j.jallcom.2011.04.139} {\bibfield  {journal} {\bibinfo  {journal} {J.
  Alloys and Compounds}\ }\textbf {\bibinfo {volume} {509}},\ \bibinfo {pages}
  {8283 } (\bibinfo {year} {2011}{\natexlab{a}})}\BibitemShut {NoStop}%
\bibitem [{\citenamefont {Gupta}\ \emph
  {et~al.}(2011{\natexlab{b}})\citenamefont {Gupta}, \citenamefont {Tayal},
  \citenamefont {Gupta}, \citenamefont {Gupta}, \citenamefont {Stahn},
  \citenamefont {Horisberger},\ and\ \citenamefont {Wildes}}]{gupta:JAP2011}%
  \BibitemOpen
  \bibfield  {author} {\bibinfo {author} {\bibfnamefont {M.}~\bibnamefont
  {Gupta}}, \bibinfo {author} {\bibfnamefont {A.}~\bibnamefont {Tayal}},
  \bibinfo {author} {\bibfnamefont {A.}~\bibnamefont {Gupta}}, \bibinfo
  {author} {\bibfnamefont {R.}~\bibnamefont {Gupta}}, \bibinfo {author}
  {\bibfnamefont {J.}~\bibnamefont {Stahn}}, \bibinfo {author} {\bibfnamefont
  {M.}~\bibnamefont {Horisberger}}, \ and\ \bibinfo {author} {\bibfnamefont
  {A.}~\bibnamefont {Wildes}},\ }\href {\doibase 10.1063/1.3671532} {\bibfield
  {journal} {\bibinfo  {journal} {J. Appl. Phys.}\ }\textbf {\bibinfo {volume}
  {110}},\ \bibinfo {eid} {123518} (\bibinfo {year}
  {2011}{\natexlab{b}})}\BibitemShut {NoStop}%
\bibitem [{\citenamefont {Blundell}\ and\ \citenamefont
  {Bland}(1992)}]{PRB92_Blundell}%
  \BibitemOpen
  \bibfield  {author} {\bibinfo {author} {\bibfnamefont {S.~J.}\ \bibnamefont
  {Blundell}}\ and\ \bibinfo {author} {\bibfnamefont {J.~A.~C.}\ \bibnamefont
  {Bland}},\ }\href {\doibase 10.1103/PhysRevB.46.3391} {\bibfield  {journal}
  {\bibinfo  {journal} {Phys. Rev. B}\ }\textbf {\bibinfo {volume} {46}},\
  \bibinfo {pages} {3391} (\bibinfo {year} {1992})}\BibitemShut {NoStop}%
\bibitem [{\citenamefont {Hope}\ \emph {et~al.}(1997)\citenamefont {Hope},
  \citenamefont {Lee}, \citenamefont {Rosenbusch}, \citenamefont {Lauhoff},
  \citenamefont {Bland}, \citenamefont {Ercole}, \citenamefont {Bucknall},
  \citenamefont {Penfold}, \citenamefont {Lauter}, \citenamefont {Lauter},\
  and\ \citenamefont {Cubitt}}]{PRB97_Hope}%
  \BibitemOpen
  \bibfield  {author} {\bibinfo {author} {\bibfnamefont {S.}~\bibnamefont
  {Hope}}, \bibinfo {author} {\bibfnamefont {J.}~\bibnamefont {Lee}}, \bibinfo
  {author} {\bibfnamefont {P.}~\bibnamefont {Rosenbusch}}, \bibinfo {author}
  {\bibfnamefont {G.}~\bibnamefont {Lauhoff}}, \bibinfo {author} {\bibfnamefont
  {J.~A.~C.}\ \bibnamefont {Bland}}, \bibinfo {author} {\bibfnamefont
  {A.}~\bibnamefont {Ercole}}, \bibinfo {author} {\bibfnamefont
  {D.}~\bibnamefont {Bucknall}}, \bibinfo {author} {\bibfnamefont
  {J.}~\bibnamefont {Penfold}}, \bibinfo {author} {\bibfnamefont {H.~J.}\
  \bibnamefont {Lauter}}, \bibinfo {author} {\bibfnamefont {V.}~\bibnamefont
  {Lauter}}, \ and\ \bibinfo {author} {\bibfnamefont {R.}~\bibnamefont
  {Cubitt}},\ }\href {\doibase 10.1103/PhysRevB.55.11422} {\bibfield  {journal}
  {\bibinfo  {journal} {Phys. Rev. B}\ }\textbf {\bibinfo {volume} {55}},\
  \bibinfo {pages} {11422} (\bibinfo {year} {1997})}\BibitemShut {NoStop}%
\bibitem [{\citenamefont {Ott}(2011)}]{SimulReflec}%
  \BibitemOpen
  \bibfield  {author} {\bibinfo {author} {\bibfnamefont {F.}~\bibnamefont
  {Ott}},\ }\href
  {\\http://www-llb.cea.fr/prism/programs/simulreflec/simulreflec.html}
  {\bibfield  {journal} {\bibinfo  {journal} {SIMULREFLEC}\ } (\bibinfo {year}
  {V1.7 2011})}\BibitemShut {NoStop}%
\bibitem [{\citenamefont {H\"aglund}\ \emph {et~al.}(1993)\citenamefont
  {H\"aglund}, \citenamefont {Fern\'andez~Guillermet}, \citenamefont
  {Grimvall},\ and\ \citenamefont {K\"orling}}]{PRB83_Hf}%
  \BibitemOpen
  \bibfield  {author} {\bibinfo {author} {\bibfnamefont {J.}~\bibnamefont
  {H\"aglund}}, \bibinfo {author} {\bibfnamefont {A.}~\bibnamefont
  {Fern\'andez~Guillermet}}, \bibinfo {author} {\bibfnamefont {G.}~\bibnamefont
  {Grimvall}}, \ and\ \bibinfo {author} {\bibfnamefont {M.}~\bibnamefont
  {K\"orling}},\ }\href {\doibase 10.1103/PhysRevB.48.11685} {\bibfield
  {journal} {\bibinfo  {journal} {Phys. Rev. B}\ }\textbf {\bibinfo {volume}
  {48}},\ \bibinfo {pages} {11685} (\bibinfo {year} {1993})}\BibitemShut
  {NoStop}%
\end{thebibliography}

%


\end{document}